\documentclass[11pt]{article}
\usepackage{theorem}
\usepackage{amsmath}
\usepackage{amsfonts}
\usepackage{amssymb}
\usepackage{euscript}
\usepackage{hyperref}
\hypersetup{colorlinks=true}
\newcommand{\R}{\mathbb{R}}
\newcommand{\C}{\mathbb{C}}

\newcommand{\F}{\mathbb{F}}
\newcommand{\Z}{\mathbb{Z}}
\newcommand{\N}{\mathbb{N}}

\newcommand{\fraP}{\mathfrak{P}}

\newcommand{\calE}{\mathcal{E}}
\newcommand{\calV}{\mathcal{V}}
\newcommand{\p}{\partial}

\newcommand{\leven}{\Lambda^{\kern -11pt 0}}
\newcommand{\lodd}{\Lambda^{\kern -11pt 1}}

{\theoremstyle{break} \theorembodyfont{\slshape} \newtheorem{defin}{Definition}[section]}
\newtheorem{propo}{Proposition}[section]
{\theoremstyle{break} \theorembodyfont{\rmfamily} \newtheorem{example}{Example}[section]}
\begin{document}
\begin{titlepage}
\begin{center}
{\large \bf Comments on non-commuting variables, \\$n$-ary algebras,\\and\\first-order differential equations}
\vskip 4em
{\large M. L\'egar\'e\;$^\ast$}
\\
\vskip 1em
Edmonton, Alberta, Canada\\
mlegare@ualberta.ca\\
\end{center}
\vskip 2em

\begin{abstract}
 
Use of certain non-commuting variables is considered in first-order differential equations. Superspace variables are discussed within the setting of first-order ordinary differential equations and $n$-ary algebras. Results on quadratic systems based on non-associative algebras are presented. A system related to  a Li\'enard system associated to an Abel's differential equation of the second kind is given as an example.  

\medskip
\noindent\sloppy{{\bf PACS} : 02.10.Hh,02.30.Hq,05.45.-a}

\medskip
\noindent\sloppy{{\bf MSC} : 34G20,17A42,17A70,34C14,15A75}

\medskip
\noindent\sloppy{{\bf Keywords} : associative products, anti-commuting variables, differential equations, non-associative algebras, non-commutative products, superspaces}

\end{abstract}
\vfill
$^\ast$ \small{For correspondence with the author, please use the email : mlegare@ualberta.ca}

\end{titlepage}

\pagebreak

\tableofcontents
\noindent
\section{Introduction}

\medskip
In this article, different aspects of certain (systems of) first-order differential and difference equations of the ordinary type are considered. Basic facts about ``$N$-ary algebras" are recalled, specific cases are studied, and extensions to different spaces, such as superspaces, are explored. It is known that first-order systems lead to higher-order systems or equations; properties and results of many systems have been for instance examined in refs \cite{Ro,Wa,KS,WE}. Systems of ordinary differential equations (o.d.e.s) are encountered in many situations, whether for example they are the Euler equations for the motion of a rigid body, they deal with relative growth rate problems (cf. for instance \cite{KS}), or they are involved in a qualitative description about the Riemannian curvature of a solution to the Ricci flow (cf. for instance ref. \cite{CK}, p.187). But mainly, the introduction of non-commuting, particularly anti-commuting / superspace, variables as well as the consideration of non-commutative products within the context of first-order differential systems is involved in this paper. An anti-commuting setting is often used in many topics, such as for example in supersymmetric systems and BRST operator based models. To our knowledge, it seems that the use of superspace variables in $N$-ary algebra inspired systems has not been much probed. Some notions used for certain systems of o.d.e.s will be briefly recalled and considered in generalizations involving non-commuting variables. An example that will be considered is a generalization of a first-order system related to an Abel's style differential equation of the second kind. The study of such extended system would be interesting. The system given as example is chosen to be simple enough to involve a quadratic contribution, and non-trivial enough to exhibit certain features such as absence of associativity and power-associativity of an associated algebra. Discretization is commented for certain systems of o.d.e.s.

In section 2, a set of basic facts and a succinct review of first-order sys\-tems related to $N$-ary algebras, as well as supervariables (anti-commuting variables) and superspaces, in addition to certain aspects of commutative products and non-associative (i.e. not necessarily associative, see \cite{Na}, p.2) algebras, will be offered. This would also allow to introduce some notation used in the following sections. Section 3 will present a generalization of first-order systems of o.d.e.s on superspaces (involving Grassmann variables) inspired by $N$-ary algebras, as well as an example of such systems of differential equations. The next section (section 4) will provide a set of results extending certain results of ref. \cite{KS} on some (quadratic) systems of o.d.e.s. 

Section 5 will consider different aspects of particular sets of o.d.e.s, such as generalizations of a system of o.d.e.s, a Li\'enard system, related to an Abel's differential equation of the second kind. A collection of certain aspects and comments, concerning mainly discretizations of o.d.e.s and systems of o.d.e.s; more specifically : systems based or inspired on $N$-ary algebras, linear systems, Lax pairs and metric equations will be offered in section 6. In conclusion, a last section will summarize results and suggest developments for possible future research. 

\medskip\noindent
\section{\textit{N}-ary Algebras, Systems of O.D.E.s and Superspaces}

\medskip
Let us consider a field $\F$ and an integer $N \geq 2$. A $N$-ary algebra over $\F$ could be defined as a $\F$-vector space $V$, endowed with a multi-linear map : $\mu : V^N \rightarrow V : (x_1 ,..., x_N) \mapsto \mu(x_1,...,x_N)$ (see \cite{Wa} for more details). Note that associativity is used for $N$-ary algebras in reference \cite{Ca}. Reference \cite{Ro} considers for a ring $R$, what are called $R$-algebras of arity $N$, as unital $R$-modules $A$ provided with a multiplication $\mu$, where generally, neither associativity nor commutativity is a property of these $R$-algebras of arity $N$. One can consult the references \cite{Ro,Wa,Ca,Ku} for more information, for instance, on definitions of subalgebras and ideals.

In what follows, particular systems are examined, and generalizations are sought and explored for systems of o.d.e.s of the following type \cite{Ro} :
\begin{defin}[Systems of o.d.e.s]
Systems of o.d.e.s based on $N$-ary algebras are defined in this article by
\begin{equation}\label{system-1}
\dot X_i = \sum_{k_1,...,k_N = 1}^n a_i^{k_1...k_N} X_{k_1} ... X_{k_N},
\end{equation}
where $i=1,...,n$, ``$\,\dot{}\,$" stands for a first-order derivative with respect to an independent variable or parameter, for instance $t$, and the coefficients $ a_i^{k_1...k_N}$ belongs to an associative, commutative unital ring $R$, which in this article, is suitably enough $\R$ or $\C$, unless otherwise mentioned. The variables $X_1, X_2, ..., X_n$ are associating but not necessarily commuting. The coefficients can be defined as structure coefficients of the $N$-ary algebra $(A,\mu)$ with respect to a certain basis $\{e^1,...,e^n\}$ of $A$.
\end{defin}
It is added that (\cite{Ro,Wa}) :
\begin{defin}[Structure Coefficients]
The structure coefficients of a $N$-ary algebra are given by
\begin{equation}
\mu(e^{i_1},...,e^{i_N}) = \sum_{i=1}^n a_i^{i_1...i_N} e^i,
\end{equation}
where $i_1,...,i_N = 1,...,n$.
\end{defin}
Thus, if one defines : $ X = \sum_{i=1}^n X_ie^i$, the above system of o.d.e.s (\ref{system-1}) can be written as :
\begin{equation}
\dot X = \mu(X,...,X)
\end{equation}
It is worth noting a few main results related  to such systems (\ref{system-1}). A first result is that for certain specific settings, systems with higher degree ($N > 2$) polynomial terms on the r.h.s. (right-hand side) of equation (\ref{system-1}) can be transformed in $N=2$ degree polynomials, leading to quadratic systems of o.d.e.s (see \cite{KS} and \cite{Kr} prop. 2.1). In other words, the polynomial systems can be seen as some Riccati systems with, however, a larger set of dependent variables. Secondly, if $V$ and $W$ are finite-dimensional vector spaces over $\F$ (of characteristic $0$), and if $f : V \rightarrow W$ is a homogeneous polynomial of degree $N$; then one can find a multi-linear and symmetric $\mu : V^N \rightarrow W$, such that :
$\mu(X,...,X) = f(X)$, for all $X \in V$ (see \cite{Wa}, theorem 1.10, p.14). The notion of derivative (Fr\'echet) of $f$ will be used in the article and is defined as (see \cite{Wa}) :
\begin{defin}[Derivative]
 The derivative of $f$ at $X$ in the direction of $Y$, denoted $Df(X)(Y)$, is
\begin{equation}\label{frechetder}
Df(X)(Y):=\left . \frac{1}{t}[f(X+tY) - f(X)]\right |_{t=0}
\end{equation}
when $X,Y \in V$
\end{defin}
Thirdly, a homogenization can be carried in order to retrieve a suitable system of differential equations (see \cite{Wa,KS}). This could be attained by adding a new variable.

\begin{example}
As an example of reduction of the order of homogeneous polynomial terms (occurring on the r.h.s.) of the eq.(\ref{system-1}) to Riccati (second degree) polynomial terms; one can consider, with here the usual matrix product, the system for two $t$ - dependent matrix variables $X_1$ and $X_2$, with third-degree homogeneous polynomials :
\begin{equation}
\dot X_1 = X_1^2X_2 + X_2X_1^2 , \quad \dot X_2 = X_1^3
\end{equation}
However, a larger set of dependent variables will be introduced in order to arrive to a system formulated as in (\ref{system-1}) with second-degree (homogeneous) polynomials on the r.h.s.. To obtain such system, one can define : $Y_1 = X_1^2, Y_2= X_2^2, Y_3 = X_1X_2, Y_4 = X_2X_1$, for a total of 6 $t$ - dependent matrix variables, which now obey to the following system of first-order :
\begin{alignat}{2}
\dot X_1 &= Y_1X_2 +X_2Y_1  & \quad\dot X_2 &= X_1Y_1 \nonumber \\
\dot Y_1 &= Y_1Y_3 + Y_3Y_1 +Y_1Y_4 + Y_4Y_1 & \quad\dot Y_2 &= Y_4Y_1 + Y_1Y_3 \\
\dot Y_3 &=Y_1Y_2+Y_4Y_3 +Y_1^2 &\quad \dot Y_4 &=Y_1^2 +Y_4Y_3 + Y_2Y_1\nonumber
\end{alignat}
It can be seen that for an initial set of $n$ matrices (e.g. $X_i$), involved in a system (\ref{system-1}) of $n$ o.d.e.s, one could consider in order to reduce the degree of the homogeneous polynomial terms in (\ref{system-1}) from degree $N$ to degree $2$, a set of (e.g. possibly $(n)^{N-1}$) additional $t$ - dependent matrix variables (e.g. $Y_j$). 
\end{example}
$\overline{\quad\quad}$
 
\smallskip
In view of probing systems generalized to involve anti-commuting variables (more specifically Grassmann variables), let us introduce a few relevant notions. Let us define following ref. \cite{Rod} a super vector space $\calV$ :  
\begin{defin}[Super Vector Space]
 A super vector space $\calV$ is a vector space with a choice of 2 subspaces labelled $\calV_0$ and $\calV_1$ such that : $\calV = \calV_0  \oplus \calV_1$. The elements of $\calV_0$ are identified as ``even", of $\Z_2$-degree $0$, that is : $v_0\in \calV_0$ has $\Z_2$-degree $|v_0| =0$, and the elements of $\calV_1$ are ``odd", of $\Z_2$-degree $1$, thus $v_1 \in\calV_1$ has $\Z_2$-degree $|v_1| = 1$.
\end{defin}
In this article, a super-commutative algebra, denoted $\Lambda_L$, for $L \in \N$, stands for the Grassmann algebra over $\R$ with $L$ generators. When $ L \rightarrow \infty$, $\Lambda_\infty$ will indicate the super-commutative algebra generated by an infinite set of (discrete) elements or generators (see ref. \cite{Rod} for details).

One can then introduce a superspace of $(p,q)$ dimensionality. Choosing $\Lambda_L = \leven_L \oplus \lodd_L$, with even subspace $\leven_L$ and odd subspace $\lodd_L$, as super-commutative algebra, a flat $(p,q)$-dimensional superspace $\Lambda^{p,q}_L$ can be defined \cite{Rod} :
\begin{defin}[Flat $(p,q)$-dimensional Superspace]
A flat $(p,q)$-dimensional superspace, denoted $\Lambda^{p,q}_L$, can be defined as
\begin{equation}
\Lambda^{p,q} = \underbrace{\leven_L \times ...\times \leven_L}_{\text{$p$ copies}}\times \underbrace{\lodd_L \times ...\times \lodd_L}_{\text{$q$ copies}}
\end{equation}
with ``even" dimension $p$, and ``odd" dimension $q$.
\end{defin}
An element of $\Lambda^{p,q}_L$ will be written as : $(x^1,...,x^p,\xi^1,...,\xi^q)$ (see ref. \cite{Rod} for details).

(Non-associative) algebras which are not associative could for instance be used in the case of quadratic polynomial terms on the r.h.s. of equation (\ref{system-1}), but let us add that (associative) $N$-products will often be considered in what follows. It is well known that a (non-associative) $2$ -ary (or binary) algebra $(V,\mu)$ will be called associative if for (see for example refs \cite{Wa,Ca}) : $\mu : V^2 \rightarrow V : (x_1,x_2) \mapsto \mu(x_1,x_2)$, where $x_i \in V$, one verifies that :
\begin{equation}
\mu(\mu(x_1,x_2),x_3) = \mu(x_1,\mu(x_2,x_3))
\end{equation}
for all $x_i \in V, i=1,2,3$. (See for example ref. \cite{Ca} for associativity with multi-linear inner compositions.)
Details and other algebra related definitions can be found for instance in refs \cite{Wa,Ca}. As for a commutative property, ref. \cite{Wa} calls a $N$-ary algebra $(V,\mu)$ commutative if $\mu(x_1,...,x_N) = \mu(x_{\pi(1)},...,x_{\pi(N)})$ for all $x_1,...x_N \in V$, and every permutation $\pi$ of the integers $1,...,N$.  

\medskip\noindent
\section{Systems of First-Order O.D.E.s}

\medskip
A simple system, which is a specific Li\'enard system, that will be generalized in different manners below, is here proposed :
\begin{equation}\label{relatedabel}
\dot x = x + \xi, \quad \dot \xi = x^2,
\end{equation}
where $x$ and $\xi$ are $\R$-valued functions of a real variable ``$t$". It will be referred to as Li\'enard (or Abel) system below. It can be mentioned that this Li\'enard system has no periodic solutions using the Bendixson - Dulac theorem. One also sees that : $\ddot x = \dot x +\dot \xi$, leads to the single non-linear second-order o.d.e. (which seems not of Painlev\'e type) :
\begin{equation}\label{abel2nd}
\ddot x - \dot x - x^2 =0
\end{equation}

Upon the transformation : $w(x) = \dot x = \frac{dx}{dt}$, it is found that (\ref{abel2nd}) becomes a first-order o.d.e. : an Abel o.d.e. of the second kind :
\begin{equation}
ww' - w = x^2,
\end{equation}
where $w=w(x), w' = \frac{dw}{dx}$. Solutions to certain Abel's o.d.e. of the second kind; typically of the type (canonical form) : $ww' - w = f(x)$, can be found, for instance, in ref. \cite{PZ}. One may consult also the ref. \cite{PZS}, and references therein, for canonical form solutions. Let us mention that this system is a particular case of the generalized problem of Briot and Bouquet : $\frac{dw}{dx} = \frac{g(x,w)}{h(x,w)}$ (see \cite{In}), where $\frac{dw}{dx} = \frac{w+x^2}{w}$. The transformation $w=\frac{1}{u}$ leads to : $u' + x^2 u^3 + u^2 = 0$, an Abel's equation of the first kind.

Let us consider the system (\ref{system-1}); allowing $X_1,...,X_n$ to be non-commuting variables belonging to a space such as $\mathcal{M}(p,\R)$, that is, $p\times p$ matrices with $\R$-valued elements. One can arrive to systems of first-order o.d.e.s such as :
\begin{equation}\label{narycirc}
\dot X_i = \sum_{k_1,...,k_N=1}^n a_i^{k_1...k_N} X_{k_1} \circ ... \circ X_{k_N},
\end{equation}
with the (structure) coefficients $a_i^{k_1...k_N}$ leading to a $N$-ary algebra. Here, the notation $\circ$ stands for a compatible deformed product between the (matrix) variables $X_i, i=1,...,n$; the product can however be chosen as the ordinary matrix product. For instance, one can consider a product as in example 1 of  ref. \cite{CGM} for $\mathcal{M}(2,\R)$, and naturally, generalizations to $\mathcal{M}(n,\R)$ deformed products could be attempted as well. One can consult ref. \cite{OS1,OS2} for information on and a classification of deformed compatible associative products. For the variables $X_i, i=1,...,n$, belonging to $\mathcal{M}(p,\R)$, one could expect from the system (\ref{narycirc}) a set of $n p^2$ o.d.e.s of the first-order for the $n p^2$ elements of those $n$ matrices. They could possibly be reconsidered as a new $N$-ary algebra related system for $n p^2$ variables, and suitable solution methods attempted there. The indices could then vary up to $n p^2$ variables, which depend on the parameter $t$.

 Another generalization is to introduce anti-commuting variables (here Grassmann variables). In this situation, the system (\ref{system-1}) or (\ref{narycirc}) could be transformed , or set (without superfields, see \cite{Rod}) such that :
\begin{equation}\label{supersystem}
\dot X_i = \sum_{k_1...k_N=1}^{p+q} f_i^{k_1...k_N} X_{k_1}\circ ...\circ X_{k_N}
\end{equation}
where $X \in \Lambda_L^{p,q}$, that is : $X_1,...,X_p \in \leven_L$, and $X_{p+1},...,X_{p+q} \in \lodd_L$. Hence : $X_i = \sum\limits_{\underline\lambda\; \text{even}} x_{i \underline \lambda} \beta_{[\underline\lambda]}$ for $i = 1,...,p$, and $X_i = \sum\limits_{\underline\lambda\; \text{odd}} \xi_{i \underline \lambda} \beta_{[\underline\lambda]}$ for $i = p+1,...,p+q$, with the generators (see notation in \cite{Rod}) : $\beta_{[\emptyset]}=1, \beta_{[1]}, \beta_{[2]},...,\beta_{[L]}$, where $\underline\lambda$ is defined as a multi-index : $\underline \lambda = \lambda_1\cdots\lambda_k$, with $1 \leq \lambda_1 \leq \cdots \leq \lambda_k \leq L$. The variables $x_{i\underline\lambda}$ and $\xi_{j\underline\lambda}$ are set as real-valued functions of $t$. Here, the coefficients $f_i^{k_1...k_N}$ could be chosen to belong to $\Lambda_L$, and for instance, either to $\leven_L$ or $\lodd_L$, in order to agree with the parity, even or odd, of the dotted variable on the l.h.s. (left-hand side) of the equation (\ref{supersystem}). The coefficients $f_i^{k_1...k_N}$ are not structure coefficients of a $N$-ary algebra as defined in section 2. Each function $X_i$ above could be respectively seen as an even or odd projection of an element belonging to a sheaf of supercommuting rings over an open set $U$ of $\R$ (see for instance \cite{Va}). 

\begin{example}
For example, one can set $L=2, N=2$ and write the above system (\ref{supersystem}) on $\Lambda_2^{2,2}$ with :
\begin{alignat}{2}
X_1 &= x^1_0 + x^1_{12} \beta_{[1]}\beta_{[2]} &\quad X_2 &= x^2_0 + x^2_{12} \beta_{[1]}\beta_{[2]}\notag\\
X_3 &=\xi_1^1 \beta_{[1]} + \xi_2^1\beta_{[2]} &\quad X_4 &=\xi_1^2 \beta_{[1]} + \xi_2^2\beta_{[2]}
\end{alignat}
The coefficients $f_i^{k_1k_2}$ could take their value(s) in $\Lambda_2$ as well. Thus, since $X_1$ is even (that is, belongs to $\leven_2$), one can set : $f_1^{k_1k_2}$ to belong to $\leven_2$  , if $k_1,k_2 \in \{1,2\}$; $f_1^{k_1k_2} \in \lodd_2$, if $k_1 \in \{1,2\}$ and $k_2 \in \{3,4\}$, or if  $k_2 \in \{1,2\}$ and $k_1 \in \{3,4\}$; and $f_1^{k_1k_2} \in \leven_2$, if $k_1,k_2 \in \{3,4\}$. Similar appropriate definitions would hold for $f_2^{k_1k_2}$.
\end{example}
\noindent
$\overline{\quad\quad}$

\smallskip
 A system of the type (\ref{supersystem}) would generally lead to a set of coupled o.d.e.s associated to different basis terms $\beta_{[\underline\lambda]}$. Non-linear systems could occur with the core (or ``body" \cite{Rod,BD}, that is, the contribution associated to $\beta_{[\emptyset]}=1$) dependent variables, since : $\beta_{[\emptyset]} \beta_{[\emptyset]} = \beta_{[\emptyset]}, \beta_{[\emptyset]} \beta_{[\underline \lambda]} = \beta_{[\underline \lambda]}$, and $\beta_{[\underline\lambda_1]}\beta_{[\underline\lambda_2]} \neq \beta_{[\underline\lambda_1]}\;\text{or}\;\beta_{[\underline\lambda_2]}$, if the multi-indices $\underline\lambda_1$ and $\underline\lambda_2$ are different from $\emptyset$. Solving at the core level, and then searching for solutions to variables related to ascending degrees of $\beta_{[\underline\lambda]}$, using solutions to variables found at lower degrees, could involve (coupled) linear differential equations. The $\Lambda_\infty$  case can also be considered. In this situation, $X_i \in \leven_\infty$, if $i =1,...,p$, $X_i \in \lodd_\infty$, if $i=p+1,...,p+q$, and the coefficients $f_i^{k_1...k_N} \in \leven_\infty \;\text{or} \; \lodd_\infty$, in a way to insure certain properties of the system, such as parity : even or odd. Super matrices could also be used as dependent variables (e.g. $X_i$), generalizing the ordinary case  discussed previously. A usual product of super matrices can be chosen, or modifications or such product along the lines given in refs \cite{CGM,OS1}, if available (i.e. existing).

Back to the Abel related system (\ref{relatedabel}) encountered above, a generalization with anti-commuting variables can be thought of as :
\begin{defin}[Generalized Li\'enard System]
A generalization of the Li\'enard (or Abel) system (\ref{relatedabel} ) presented above has the form
\begin{equation}\label{genabel}
\dot x = x+\gamma \xi, \qquad \dot \xi  = \alpha x^2
\end{equation}
where $x \in \leven_\infty, \xi \in \lodd_\infty$ play the role of dependent variables of $t$, $\dot x$ and $\dot \xi$ express derivatives of first-order with respect to $t$, and where $\alpha, \gamma \in \lodd_\infty$ are constants. 
\end{defin}
It is not a standard supersymmetric construction. It is mentioned that the system : $\dot x = \gamma \xi, \dot \xi = \alpha x^2$, does not present certain characteristics later discussed. It follows that for $x$ : 
\begin{equation}
\ddot x = \dot x + (\gamma\alpha) x^2
\end{equation}
where $(\gamma\alpha) \in \leven_\infty$ and $\ddot x$ is the second-order derivative of $x$ with respect to $t$. A formulation closer to the one presented in section 2 can be provided when one uses for instance : $X = \sum\limits_{i=1}^{p+q} X_i E^i$, for a linearly independent set $\{E^i= (\delta_1^i,...,\delta_{p+q}^i), i = 1,...,p+q\}$, where $X_i \in \leven_\infty \,(i=1,...,p)\text{or}\; \lodd_\infty (i = p+1,...,p+q)$. For $V = \Lambda^{p,q}_\infty = (\,\leven_\infty)^p \times (\,\lodd_\infty)^q$, $\mu : V^N \rightarrow V$, can be related to :
 \begin{equation}
\mu(E^{k_1},..., E^{k_N}) = \sum_{i=1}^{p+q} b_i^{k_1 ... k_N} E^i
\end{equation}
where $k_1, ..., k_N = 1, ... , p+q$, and the coefficients $b_i^{k_1 ... k_N} \in \Lambda_\infty$.

Therefore, an equation : $\dot X = \mu(X,...,X)$ can be written with extended multilinearity, similar to equation (\ref{supersystem}) up to redefinitions of constants, where $X$ is defined as above. When $N=2, p=1, q=1$, that is : $X = X_1 E^1+ X_2E^2$, with $X_1$ and $X_2$ chosen respectively as $X_1 = x \in \leven_\infty$ and $X_2=\xi \in \lodd_\infty$, it is derived that :
 \begin{equation}
\sum_{i=1}^2  \dot X_i E^i = \sum_{k_1,k_2 =1}^2 X_{k_1}X_{k_2}\mu(E^{k_1},E^{k_2})
\end{equation}
and thus :
\begin{equation}
\dot x E^1 + \dot \xi E^2 = \sum_{k_1,k_2=1}^2 \sum_{i=1}^2  X_{k_1}X_{k_2}b_i^{k_1 k_2} E^i
\end{equation}
For an appropriate choice of coefficients $b_i^{k_1 k_2}$ ($b_1^{k_1k_2} = 0$, for any $k_1,k_2$, $b_2^{11}=\alpha$, otherwise $0$), one can retrieve the homogeneous second-degree contribution of the generalized Li\'enard system (\ref{genabel}).

\medskip\noindent
\section{Results for Systems of First-Order O.D.E.s and anti\--com\-mu\-ting setting}

\medskip
It has already been mentioned that systems of o.d.e.s involving non-com\-muting variables in a homogeneous polynomial term of degree larger than  2 can be rewritten as homogeneous second-degree polynomial terms involving a larger set of variables. In the fashion of refs \cite{Wa,KS,Kr}, let us consider non-associative algebras constructions with commutative products. More specifically, let us consider the homogeneous second-degree polynomial system as a particular case ($N=2$) of equation (\ref{system-1}) (without Grassmann (anti-commuting) variables here) :
\begin{equation}\label{quadsystem}
\dot X_i = \sum_{k_1,k_2= 1}^n a_i^{k_1k_2} X_{k_1} \ast X_{k_2}, \quad \text{or} \quad \dot X = \sum_{i=1}^n \sum_{k_1k_2=1}^n a_i^{k_1k_2} X_{k_1} \ast X_{k_2} e^i
\end{equation}
where $i,k_1,k_2 = 1, ..., n$, and $\ast$ is a possibly non-commutative but associative product. Let us mention that the system (\ref{quadsystem}) can be cast into another system : $\dot X = X \circ X$, where the notation $\circ$ stands for a non-associative commutative product (see \cite{Wa}). Using the derivative defined in equation (\ref{frechetder}) where $f(X)$ is the polynomial found on the r.h.s. of equation (\ref{quadsystem}), one obtains that :
\begin{equation}
P(X) = \frac{1}{2!} D^2f(X) (X,X) = X \circ X,
\end{equation}
where the product $\circ$ is non-associative, but symmetrized (commutative). Extending a previous mention in section 2, one can rewrite a system of o.d.e.s $\dot X = \mathcal{P}(X)$, where $\mathcal{P}(X)$ is a polynomial of degree $N$ in $X$ (not necessarily homogeneous), as a larger system : $\dot{\tilde X} = \tilde{\mathcal{P}}(\tilde X)$, where $\tilde{\mathcal{P}}(\tilde X)$ is a second-degree polynomial, with $\tilde X$ representing a larger set of dependent variables (\cite{Wa,Kr}).

\begin{example}
As an example of such quadratic system (but now in superspace), let us consider the generalized Li\'enard system (\ref{genabel}), and rewrite it as : 
\begin{equation}\label{superquad}
\dot X  = C + T X + Q(X)
\end{equation}
in analogy with the formulation found in \cite{KS}. In a superspace setting, define $X = \bmatrix x \\\xi \endbmatrix, C = \bmatrix 0 \\ 0 \endbmatrix, T = \bmatrix 1 & \gamma \\0 & 0 \endbmatrix$, and $Q(X) = Q(x,\xi) = \bmatrix 0 \\ \alpha x^2 \endbmatrix$ ; hence one can set for the quadratic term in equation (\ref{quadsystem})  with the usual product selected for the product $\ast$ : $a_1^{k_1k_2} = 0, \forall k_1,k_2, a_2^{11} = \alpha$, otherwise $a_2^{k_1k_2} = 0$. 
\end{example}
\noindent
$\overline{\quad\quad}$

\smallskip
One can also carry out a ``homogenization" (see \cite{Wa,KS}), by bringing a variable $u \in \leven_\infty$ such that : $\dot u = 0$, which implies that $u$ is a constant. The following extended system can then be provided :
$\dot X  = \tilde Q(\tilde X) = u^2C + uT X + Q(X), \quad \dot u = 0$, with $\tilde X = \bmatrix X \\u \endbmatrix$.
The latter (extended) system can be shortly written as :
\begin{equation}\label{quadhomogen}
\dot{\tilde X} = \bmatrix \tilde Q(\tilde X) \\ 0 \endbmatrix, \quad \text{or} \quad \bmatrix \dot X \\ \dot u \endbmatrix = \bmatrix \tilde Q(\tilde X) \\0 \endbmatrix
\end{equation}
Solutions to the system (\ref{quadhomogen}) for $X,u$, or $\tilde X = \bmatrix X \\u \endbmatrix = \bmatrix x \\ \xi \\ u \endbmatrix$, lead to solutions for $X$ by using the subset $u=1$. It can be verified that : $Q(\lambda X) = \lambda^2 Q(X), \tilde Q(\lambda \tilde X) = \lambda^2 \tilde Q(\tilde X)$, where $\lambda \in \R$. One can then obtain, as indicated above using (Fr\'echet) derivatives, the (bilinear map) product : $\beta(X,Y) = \frac{1}{2} D^2Q(0)(X,Y)$ : 
\begin{equation}
\beta(X,Y) = \frac{1}{2}\left [Q(X+Y) - Q(X) - Q(Y) \right ]  = \bmatrix 0 \\ \frac{\alpha}{2}(xy+yx) \endbmatrix = \bmatrix 0 \\\alpha xy\endbmatrix 
\end{equation}
with $X = \bmatrix x \\ \xi \endbmatrix , Y = \bmatrix y \\ \chi \endbmatrix$, $x,y \in \leven_\infty, \xi,\chi \in \lodd_\infty$.
Therefore $\beta(X,X) = X \circ X = \bmatrix 0 \\ \alpha x^2 \endbmatrix$,\newline  and the extended product follows as :\newline $\tilde \beta(\tilde X, \tilde Y) = \tilde X \circ \tilde Y =  \frac{1}{2}\left [\tilde Q(\tilde X + \tilde Y) - \tilde Q(\tilde X) - \tilde Q(\tilde Y) \right ]$. Explicitly, one has :
\begin{equation}
\tilde \beta(\tilde X,\tilde Y) = \tilde X \circ \tilde Y = \bmatrix uvC + \frac{1}{2}(uTY + vTX) + \beta(X,Y) \\ 0 \endbmatrix
\end{equation}

From these, one can rewrite the homogenized system (\ref{quadhomogen}) as : 
\begin{equation}
\dot{\tilde X} = \tilde X \circ \tilde X = \beta(\tilde X,\tilde X) = \bmatrix u^2C +uTX + \bmatrix 0 \\ \alpha x^2 \endbmatrix \\ 0 \endbmatrix = \bmatrix ux + u\gamma \xi \\ \alpha x^2 \\0 \endbmatrix
\end{equation}
where $\alpha, \gamma \in \lodd_\infty$ are constants, $u, x \in \leven_\infty$, and $\xi \in \lodd_\infty$. It has been found that the product $\circ$ : $\tilde X \circ \tilde Y$, is commutative (symmetrized) but neither associative, nor power-associative (see \cite{Wa, Sc} for definitions and examples). Power series style solutions are then even less of interest, but possible and tedious to express (see \cite{Wa,KS}). In fact, a series could be considered (cf. \cite{KS}) :
\begin{subequations}
\begin{align}
\tilde X(t) &= \tilde X(0) + \tilde X^{(1)}(0) t + \tilde X^{(2)}(0) \frac{t^2}{2!} + \cdots\\
\tilde X(t) &= \tilde X(0) + \tilde X^2(0) t + \tilde X^3(0) t^2 + (2\tilde X^4(0) + \tilde X^2(0) \circ \tilde X^2(0)) \frac{t^3}{3} + \cdots\label{series}
\end{align}
\end{subequations}
with initial values $\tilde X(0)$, and $\tilde X^l(t)$ defined as : $\tilde X^l(t) = \tilde X(t) \circ \tilde X^{l-1}(t) $ for $l = 2,3,...$.  
However, a solution to the simpler system : $\dot x = u\gamma \xi, \dot \xi = \alpha x^2, \dot u = 0$, can be obtained in ``closed" form. The associated algebra is power-associative; one computes the following solution using the series (\ref{series}) above, truncated since $\alpha, \gamma \in \lodd_\infty$ :
\begin{subequations}
\begin{align}
x(t) &= x(0) +u(0) \gamma \xi(0) t +\frac{1}{2}u(0)\gamma\alpha x^2(0) t^2\\
\xi(t) &= \xi(0) + \alpha x^2(0) t + \alpha x(0) u(0) \gamma \xi(0) t^2\\
u(t) &= u(0)
\end{align}
\end{subequations}
with initial values $x(0), \xi(0)$ and $u(0)$.
It is noted that systems (\ref{superquad}) with variables in $\Lambda^{p,q}_L$, where $L$ is finite, can be considered in a similar fashion. One can also mention that the system $\dot x= u(x + \xi), \dot \xi = x^2, \dot u = 0$ where $x, \xi, u$ are real-valued functions, can be considered with the series (\ref{series}). This could be extended to attempt deriving a series solution of the equation : $\ddot x - \dot x - f(x) =0$, encountered in section 3, with $f(x)$ as a real-valued polynomial of degree higher than 2. A first step could be to rewrite the second-order equation via the system : $\dot x = x + \xi, \dot \xi = f(x)$. Convergence of derived series would have to be considered.

In the above spirit, an extension to anti-commuting variables of certain definitions and results of ref. \cite{KS} can be attempted below. According to \cite{KS}, the following can be defined.
\begin{defin}[Automorphism]
An automorphism of an algebra $A = (V,\beta)$, would be defined as an invertible linear transformation $\phi \in GL(V)$, such that : $\phi\beta(X,Y) = \beta(\phi X, \phi Y)$, for all $X, Y \in A$.
\end{defin}
Also, 
\begin{defin}[Derivation]
A derivation of an algebra $A = (V,\beta)$ would be defined as a linear transformation $D : A \rightarrow A$, obeying to the product rule : $ D \beta(X,Y) = \beta(DX,Y) + \beta(X,DY)$, for all $X, Y \in A$.
\end{defin}
As a generalization to superspace of such notions, one could imagine setting $V = \Lambda^{p,q}_\infty$ with basis $\{\underbrace{e^1, ... , e^p}_{even},\underbrace{e^{p+1}, ... ,e^{p+q}}_{odd}\}$, \newline $\beta(X,Y) = \frac{1}{2}\left [Q(X+Y) - Q(X) -Q(Y) \right ]$, and identify $GL(V)$ as the super Lie group $GL(p,q, \Lambda_\infty)$ (with even matrices as diagonal blocks) \cite{Ber}. Analogously (see ref. \cite{KS}), an automorphism of a ``vector" $E(X) = C + TX + \beta(X,X)$ is set as an invertible linear transformation $\phi \in GL(V)$, such that : $E(\phi X) = \phi E(X)$, for all $X \in V$.  A derivation of a ``vector" $E(X)$  is then chosen as a linear transformation : $D : V \rightarrow V$, obeying to : $DE(X) = E'(X) (DX)$, where $E'(X)$ is a linearization of $E(X)$, for all $X \in V$. With respect to a previous structure introduced, one could use specifically : $X = \bmatrix x \\ \xi \endbmatrix \in \Lambda^{1,1}_{L \; \text{or} \; \infty}$ with $E(X) = TX +Q(X) = \bmatrix 1 & \gamma \\0 & 0 \endbmatrix \bmatrix x \\ \xi \endbmatrix + \bmatrix 0 \\ \alpha x^2 \endbmatrix$.

However, let us suppose that $X \in \Lambda^{p,q}_\infty$ and that : $E(X) = TX + \beta(X,X) = TX + X\circ X$. It can be deduced that the condition  $\phi E(X) = E(\phi X)$, for any $X \in \Lambda^{p,q}_\infty$, implies that : $\phi TX = T\phi X$, or $\phi T = T \phi$, and $\phi(X \circ X) = \phi X \circ \phi X$. It can hence be said that $\phi$ is an automorphism of $E(X)$ if $\phi$ is an automorphism of $A = (\Lambda^{p,q}_\infty,\beta)$, such that $\phi T = T \phi$. Similarly, $D$ is a derivation of $E(X) = TX + \beta(X,X) = TX + X\circ X$, if $D$ is a derivation of $A = (\Lambda^{p,q}_\infty, \beta)$, such that $TD = DT$. One can then propose the following results.

Firstly (related to proposition 5.3 of \cite{KS}) :
\begin{propo}
Let $G$ belongs to a ``superalgebra" (called $gl(p,q,\Lambda_\infty)$) of the super Lie group $GL(p,q,\Lambda_\infty)$ of linear transformations of $\Lambda^{p,q}_\infty$, where it is assumed that $G$ is such that : $e^{tG} \in GL(p,q,\Lambda_\infty)$, where  $e^{tG} = Id + \sum\limits_{i=1}^\infty \frac{t^i}{i!} G^i$, for $t \in \R$, and let $G$ be such that $e^{tG}$ is an automorphism of $E$ (here defined as $E(X) = TX + X\circ X)$. Then $e^{tG}P$, with $P$ an element of $A$ is a solution to : $\dot X  = E(X)$, if and only if $GP =E(P)$.
\end{propo}

This result follows from : $\frac{d}{dt} (e^{tG}) P  = E(e^{tG} P)$, since $e^{tG}$ belongs to the automorphisms of $E$.
$\square$

Secondly, some results with respect to idempotents can also be extended (c.f. lemma 3.2 of \cite{KS}). 
\begin{propo}
Let $\varepsilon \in A = (\Lambda^{p,q}_\infty,\beta)$ be defined as an idempotent element of $A$ if $\varepsilon^2 = \varepsilon \circ \varepsilon = \beta(\varepsilon, \varepsilon) = \varepsilon \neq 0$. Then if one considers the system : $\dot X = X \circ X = \beta(X,X)$, where the product $\circ$ is non-associative but commutative, it can be derived that : $F_t(\varepsilon) = \frac{1}{1-t} \varepsilon$, is an (unbounded) solution to $\dot X = \beta(X,X) = X \circ X$, which blows up in finite $t$.
\end{propo}
One puts $X = G(t) = g(t) \varepsilon$, with $g : \R \rightarrow \R$ and $\varepsilon \in \Lambda^{p,q}_\infty$. Therefore : $\beta(G(t),G(t)) = g^2(t) \beta(\varepsilon,\varepsilon) = g^2(t) \varepsilon$. The system $\dot X  = X \circ X$ is verified if one has the equation $\dot g(t) = g^2(t)$ satisfied. Let us impose the initial condition $g(0) = 1$, then $G(t) = \frac{1}{1-t} \varepsilon$, as a solution. $\square$

If now one supposes that $P\neq 0 \in (\Lambda^{p,q}_\infty, \beta)$, and obeys to $P^2  = P \circ P= a P$, where $a \in \R \setminus 0$. Then (as a modification of corollary 3.3 of \cite{KS}), it is found that : $X(t) = \frac{1}{1-at} P$, is also a solution of the system $\dot X = X \circ X$ above. This solution is unbounded and goes to infinity in a finite positive $t$ for $a>0$, and in a finite negative $t$ for $a<0$. One also looks at the case when (see proposition 3.4 of \cite{KS}) $A=(\Lambda^{p,q}_\infty,\beta)$ has a nontrivial (i.e. $\neq 0$) idempotent $\varepsilon$, then the origin $X=0$ of $A$ might have a special role as equilibrium point for the system $\dot X = X \circ X = \beta(X,X)$. Note that if $P \neq 0$ is such that $P \circ P = 0$ (equilibrium point for the latter system), then $P$ is also nilpotent but with respect to the product $\circ$. Let us add that other results of ref. \cite{KS} can also be considered for formulation in superspaces (for example $\Lambda^{p,q}_\infty$).

\medskip\noindent
\section{Symmetries, O.D.E.s and Superspaces}

\medskip
This section primarily explores a notion of ``symmetries" for o.d.e.s associated with superspaces. It is an attempt to consider certain aspects of the work presented in refs \cite{WE,MS1,MS2} in a superspace setting. As an example, a particular system, the simple generalization to superspace of a Li\'enard system (equation  (\ref{genabel})) encountered in previous sections of this article, will be discussed. It is seen in \cite{WE,MS1,MS2} that similarities to partial differential equation properties, for instance symmetries,  can be found in certain systems of o.d.e.s on free associative algebras. Some notions used in those papers are adapted in this section. Let us first introduce \cite{MS1} an ``infinitesimal (symmetry) generator".
\begin{defin}[Infinitesimal Generator] 
$\mathsf{G}$ of a differential system $M_t = \mathsf{F}$ on an algebra, denoted  $\fraP$, of polynomials in $\Lambda^{p,q}_\infty$ variables, is said to be an infinitesimal generator when $M_\tau = \mathsf{G}$, and $M_t = \mathsf{F}$ commute, hence if : $\frac{d \mathsf{G}}{dt} = \frac{d \mathsf{F}}{d\tau}$. 
\end{defin}
Therefore, with the notation $D_tM  = \mathsf{F}$ and $D_\tau M = \mathsf{G}$, the derivations $D_t$ and $D_\tau$ would satisfy : $D_tD_\tau = D_\tau D_t$, acting on $\fraP$.

 The (Fr\'echet) derivative (see also (\ref{frechetder})) can be given as : 
\begin{equation}\label{frechetdersuperspace}
\frac{d}{d \epsilon}f(u_i + \epsilon \delta u_i) |_{\epsilon = 0} = f_\ast(\delta u_i),
\end{equation}
where $i = 1, ..., p+q, u_i \in \leven_\infty \; \text{or}\; \lodd_\infty$.

  For example, one could have the variables $(x,\xi) \in \Lambda^{1,1}_\infty$, i.e. $u_1 = x \in \leven_\infty, u_2 = \xi \in \lodd_\infty$, with the system :

\begin{subequations}
\begin{align}
x_t &= \mathsf{F}_1(x,\xi) = a_0 + \sum_{k=1}^\infty a_k x^k + \sum_{k=1}^\infty b_k x^k \xi  + c \xi\\
\xi_t &= \mathsf{F}_2(x,\xi) = \alpha_0 + \sum_{k=1}^\infty \alpha_k x^k + \sum_{k=1}^\infty \beta_k x^k \xi  + \gamma \xi 
\end{align}
\end{subequations}
where $\mathsf{F}_1(x,\xi)$ and $\mathsf{F}_2(x,\xi)$ are set as polynomials in $x$ and $\xi$, with constant coefficients $a_0,a_k,\beta_k,\gamma \in \leven_\infty$, and $\alpha_0,\alpha_k,b_k,c \in \lodd_\infty$, and $x_t, \xi_t$ stands for respective derivatives of $x$ and $\xi$ with respect to $t$. 

\begin{example}[Infinitesimal generator]
In particular, a generalized Li\'enard system could be expressed as (see equation (\ref{genabel})) :
\begin{equation}\label{system-F}
x_t = \mathsf{F}_1(x,\xi) =  x+ e\xi, \quad \xi_t = \mathsf{F}_2(x,\xi) = \alpha x^2
\end{equation}
with nonzero $e,\alpha \in \lodd_\infty$. (Note that the role of $\gamma$ in (\ref{genabel}) has been relayed to $e$ for notational purposes.) Let us define  for $M_\tau = \mathsf{G}$, with $M = (x,\xi)^T$, the system :
\begin{subequations}\label{symrel}
\begin{align}
x_\tau &= \mathsf{G}_1 = a_0 + a_1 x+ a_2 x^2 + a_3 x^3+ b_1 x\xi + b_2 x^2 \xi +b_3 x^3 \xi + c \xi  \\
\xi_\tau & = \mathsf{G}_2 = \alpha_0 + \alpha_1 x + \alpha_2 x^2 + \alpha_3 x^3 + \beta_1 x\xi + \beta_2 x^2 \xi + \beta_3 x^3 \xi + \gamma \xi
\end{align}
\end{subequations}
where  $a_0,a_1,a_2,a_3,\beta_1,\beta_2,\beta_3,\gamma \in \leven_\infty$, and $b_1,b_2,b_3,c,\alpha_0,\alpha_1,\alpha_2,\alpha_3 \in \lodd_\infty$
Imposing the commutativity of the derivations $D_t$ and $D_\tau$ acting (on the polynomials) on $\Lambda^{1,1}_\infty$, involves 16 unknowns ($\in \leven_\infty$ or $\,\lodd_\infty$) with a set of 20 equations. Given the arbitrariness of the nonzero constants $e$ and $\alpha$, a solution set can be formulated as follows :
\begin{alignat}{4}  
a_0 &= -e\alpha_0 & \quad a_1 & \in \leven_\infty & \quad a_2 & = \frac{e\gamma\alpha}{2} &\quad a_3 & = -\frac{b_1 \alpha}{2}  \\
b_1 &\nonumber \in \lodd_\infty& \quad b_2 &= \frac{b_1 \alpha e}{2}&\quad b_3 & = 0 &\quad c & = (a_1 - \gamma) e  \\
\alpha_0 &\nonumber \in \lodd_\infty& \quad \alpha_1 &= 2 \alpha\alpha_0 e&\quad \alpha_2 & = (2a_1-\gamma)\frac{\alpha}{2} &\quad \alpha_3 & = 0 \nonumber  \\
\beta_1 &= e\gamma \alpha& \quad \beta_2 &= \alpha b_1&\quad \beta_3 & = 0 &\quad \gamma & \in \leven_\infty  \nonumber 
\end{alignat}
or as :
\begin{subequations}\label{vector-G}
\begin{align}
x_\tau &= \mathsf{G}_1 = a_1(x+e\xi) -e\alpha_0 + \gamma\left (\frac{e\alpha x^2}{2} - e\xi \right ) + b_1 \left (\frac{-\alpha}{2} x^3 + x\xi + \frac{\alpha e x^2 \xi}{2} \right ) \\
\xi_\tau & = \mathsf{G}_2 = a_1 \alpha x^2 + \alpha_0\left (1-2\alpha e x \right ) + \gamma \left (\frac{-\alpha x^2}{2} + \xi + e\alpha x \xi \right ) - b_1 \alpha x^2 \xi
\end{align}
\end{subequations}
where $a_1,b_1,\alpha_0$ and $\gamma$ are arbitrary variables belonging appropriately to $\leven_\infty$ or $\lodd_\infty$.
\end{example} 
\noindent
$\underline{\quad\quad}$

\smallskip
These ``transformations", the equations  (\ref{vector-G}), relate the variable ($x$) in $\leven_\infty$ and the variable ($\xi$) in $\lodd_\infty$. Same type of results can be obtained for the system : $x_t = e\xi, \xi_t = \alpha x^2$.

To provide a comparison, let us add, that for a non-commutative product $\circ$, seeking ``symmetries" of the form : $x_\tau = P(x,\xi), \xi_\tau = Q(x,\xi)$, where $P$ and $Q$ are polynomials of degree 3 in $x$ and $\xi$ (not superspace variables) with real coefficients leads to a trivial solution : $ x_\tau = c(x+\xi), \xi_\tau = x \circ x$.
As a side comment, let us mention that it is well known that the system (\ref{relatedabel}) where $x$ and $\xi$ are real-valued, leads to a single partial differential equation for two infinitesimal symmetry prolongation parameters (see \cite{Ol}), of the equation : $\frac{d\xi}{dx} = \frac{x^2}{x+\xi}$. Its (non-trivial) solution would possibly reveal some (ordinary) symmetries.

Let us now define analogously to refs \cite{WE,MS1}, operators of left (right) multiplication
\begin{defin}[Left and Right Multiplications]
Operators of left (right) multiplication on the space of polynomials $\fraP$ in $\Lambda^{p,q}_\infty$ variables, considered above are denoted $L_a (R_a)$. Explicitly, one has : $L_a(z) = az (R_a(z) = za)$, where $a,z \in \fraP$.
\end{defin}
The (Fr\'echet) derivative defined earlier allows to write : $D_t a  = \vec a_\ast(\vec{\mathsf{F}})$, where $\vec{\mathsf{F}}$ is the r.h.s. of the system (\ref{system-F}). Hence :
\begin{propo} Let $\vec{\mathsf{F}} = (\mathsf{F}_1= x+\xi, \mathsf{F}_2 = \alpha x^2)$, it is then that : $\mathsf{F}_{1_\ast} = (1,L_e)$ and $\mathsf{F}_{2_\ast} = (L_\alpha L_x +R_xL_\alpha,0)$. It follows that : $D_t\mathsf{G}_i = D_\tau \mathsf{F}_i, i =1,2$, that is : $D_t\vec{\mathsf{G}} = D_\tau\vec{\mathsf{F}} = \vec{\mathsf{F}}_\ast(\vec{\mathsf{G}})$, using $\vec{\mathsf{G}} = (\mathsf{G}_1,\mathsf{G}_2)$ with the components $\mathsf{G}_i$ given in (\ref{vector-G}).
\end{propo}

This can be shown using the above definition of derivative (\ref{frechetdersuperspace}). $\square$

Hamiltonian aspects with the above approach can be found in refs \cite{WE,MS1}. It is not our goal to look for and probe Hamiltonian formulations for systems considered in this paper.

\medskip\noindent
\section{Comments and Related Equations}

\medskip
In this section, a small set of short aspects on certain main topics covered in the previous sections is explored. A discretization or difference equation perspective is often added or brought. This section can be seen as overlapping the conclusion. May it be reminded that different discretization or difference equation schemes can often be suggested or selected for certain differential equations or systems; one can consult, for instance, discussions in refs \cite{Mi,BS,GRSW}. However, in what follows, a simple approach to difference setting will be considered in order to get an initial point of view.

\subsection{Difference First-Order Systems}
First, let us suggest for the general $N$-ary systems of o.d.e.s (\ref{system-1}) a difference formulation :
\begin{equation}\label{discretesystem-1}
X_i(\tau+1) = X_i(\tau) + \sum_{k_1...k_N = 1}^n h a_i^{k_1 ... k_N} X_{k_1}(\tau) \circ \cdots \circ X_{k_N}(\tau)
\end{equation}
where $h \neq 0$ is a constant, the product $\circ$ is non-associative, and the discrete variable $\tau$ takes positive integer values (1,2,...). An equilibrium solution corresponds to $X_i(\tau+1) = X_i(\tau)$, for all $\tau$ and $i=1,...,n$, which implies that $\mu(X, ...,X) = 0$, for any $\tau$. An idempotent, denoted $\varepsilon$, could be defined as $\varepsilon$ such that : $\mu(\varepsilon,..., \varepsilon) = \varepsilon$, for any $\tau$. Hence : $X_i(\tau +1) = X_i(\tau) + hX_i(\tau) = (1+h) X_i(\tau)$, for all $i=1,...,n$, where $(1+h)$ is a scale factor, leading to solutions along a line in $n$-dimensional space.

It is stressed that different products $\circ$ can be used and that a $N$-ary system can be reduced to a quadratic system with the addition of new variables. However, as in previous sections, it would be interesting to consider this discretization with superspace variables. No major change to the form of the equation (\ref{discretesystem-1}) is expected.

\subsection{Linear Systems and First-Order Systems}
Secondly, let us indicate that many quadratic systems, such as systems introduced in sections 3 and 4 (for instance, certain systems of the type (\ref{quadsystem})) can be seen as special cases of matrix Riccati equations. This means that one could have (see \cite{Wa,Re,Le}) :
\begin{equation}\label{riccati}
\dot X = X \circ A \circ X + B \circ X + X \circ C + D
\end{equation}
where $X$ belongs to the set of $p\times q$ matrices with real- or complex-valued elements, and appropriately constant matrices $A$ of dimensions $q\times p$, $B$ of dimensions $p\times p$, $C$ of dimensions $q\times q$, and $D$ of dimensions $p\times q$, with associative product $\circ$. It is recalled (see for example ref. \cite{Wa}) that a linearization of the system (\ref{riccati}) can be obtained as follows :
\begin{equation}
\bmatrix \dot u  \\ \dot v \endbmatrix = \frac{d}{dt} \bmatrix u  \\ v \endbmatrix = \bmatrix  B & D \\-A & -C \endbmatrix  \circ \bmatrix u  \\ v \endbmatrix
\end{equation}
leading to a first-order system with matrices $u$ of dimensions $p \times q$ and $v$ of dimensions $q\times q$, the latter matrix ($v$) assumed  to be invertible. Solutions $X = u \circ v^{-1}$ are then derived. For the difference case with ordinary matrix product chosen for the product $\circ$, one could define the difference operator $\Delta$ : $\Delta f(\tau) = \frac{[f(\tau +1) - f(\tau)]}{h}$, the translation operator $\calE$ : $\calE f(\tau) = f(\tau +1)$, and use the ``Leibniz" or product rule :
\begin{equation}\label{productrule}
\Delta (fg) = \Delta f \calE g + f \Delta g
\end{equation}
The discretized system : $\Delta u = b\calE u + d \calE v, \Delta v = -a\calE u -c\calE v$, would then produce the matrix difference equation : $\Delta x = x a \calE x + b \calE x + xc + d$, where $ x = u v^{-1}$. One may refer to \cite{PW} for other discretizations . (Let us note that the systems seen previously in equations (\ref{relatedabel}) or (\ref{genabel}) do not lend themselves to systems of the matrix Riccati differential type with $X = \bmatrix x \\ \xi \endbmatrix$.) Consideration of variables in superspaces would be possible, for instance by allowing (super-)matrices belonging to linear transformations acting on suitable superspaces. One notes that the system $\dot X = E(X)$ encountered above in section 4 has been given a discretization in section 6 of ref. \cite{KS} , that is, $\Delta X(\tau) = E(X(\tau))= T X(\tau) + \beta(X(\tau),X(\tau))$. 

The ``complete" integrability of a system of $N$ autonomous ordinary differential equations can follow from the existence of a suitable set of symmetries or invariants (see \cite{MS2,Ol}). For certain systems, ``complete" integrability is also linked to the existence of Lax pairs or linear systems. Discretizations of such notions have been explored in different works (see for instance refs \cite{BS,RSWa,WZG,DMH,Hie,Ha}). Indeed, some discretized formulations of Lax systems or associated linear systems (analogous to the above discretized system for the set of $u$ and $v$) (see \cite{WZG}) with discretized gauge transformations have been studied. Again, the use of different products and dependent variables with values in superspaces could be attempted.

\subsection{Certain Metric Equations}

Looking at systems of o.d.e.s associated to the evolution of the curvature operator of a solution to the Ricci flow (see ref. \cite{CK}) brings a short comment. It might have been examined already, but one could wonder for instance about equations (flows) of the following forms preserving tensorial properties, in the style of equation (\ref{system-1}) :
\begin{equation}
\frac{\p g_{ij}}{\p t} (p,t) = a_{ij}^{k_1l_1k_2l_2 ... k_Nl_N} (p,t) g_{k_1l_1}(p,t) g_{k_2l_2}(p,t) \cdots g_{k_Nl_N}(p,t)
\end{equation}
for the parametrized metric $g_{ij}$, where the point ($p$) and parameter ($t$) dependent coefficients $a_{ij}^{k_1l_1k_2l_2 ... k_Nl_N} (p,t)$ are related to suitable $N$-ary algebras. (This could be seen as an extension of a dilation flow ($\dot g  =  a g$).) A condition could be sought for self-similar solutions of the type : $g(t) = \sigma(t) \varphi^\ast_t(g(0))$, for diffeomorphisms $\varphi$  and scale factor $\sigma(t)$ (see \cite{CK,Ben}) on the metric $g$. As for (Hamilton's) Ricci flow (see for instance \cite{MMAGY} and some references therein), discretizations (differential difference equations) could be considered.

It is mentioned that many aspects in this section have not been addressed in details, further explorations of these discrete systems are not the object of this article.

\medskip\noindent
\section{Conclusion}

\medskip
In this paper, main results include a probe of $N$-ary algebra inspired systems extended to anti-commuting variables, and discussions of $N$-ary inspired systems with the use of non-commu\-tative products. More specifically, systems of ordinary differential equations based on $N$-ary algebras have been generalized to superspaces. Particular systems have been probed, such as an extension to superspace of a system, a Li\'enard system,  related to an Abel's differential equation of the second kind. Rewriting a $N$-ary inspired system for $N>2$ as a system with second-degree polynomials with supplementary variables can be of interest. Certain transformations or ``symmetries" have been defined borrowing an approach taken for o.d.e.s on free associative algebras \cite{WE,MS1,MS2}. Properties of $N$-ary inspired systems with non-associative setting have been presented following steps and results of \cite{KS}. Let us mention that $\Z_2$-graded algebras with quadratic differential systems have been studied in \cite{HK}. Finally, some comments and notes on certain aspects of a discretization or difference equations point of view have been offered. 

Future works could be directed toward more focused investigations of superspace generalizations, discretizations on superspaces,  or could consider in detail some non-commutative products with $N$-ary algebra inspired equations. Discussions of the structure of certain systems (with or without superspace setting) and their solution sets could as well be explored.



\vskip 2in


\begin{thebibliography}{99}\scriptsize

\bibitem{Ro} Helmut R\"ohrl, ``Algebras and Differential Equations", Nagoya Math. J.{ \bf 68} (1977) ,pp.  59-122.

\bibitem{Wa} Sebastian Walcher, ``Algebras and Differential Equations", Hadronic Press Inc., Palm Harbor, U.S., 1991.

\bibitem{Na} Nathan Jacobson, ``Lie algebras", Dover Publ., N.Y., 1979.

\bibitem{KS} M.K. Kinyon and A.A. Sagle, ``Quadratic Dynamical Systems and Algebras", J. Diff. Eq. {\bf 117} (1995), pp. 67-126.

\bibitem{WE} T. Wolf and O. Efimovskaya, ``On Integrability of the Kontsevich Non-Abelian ODE System", Lett. Math. Phys. {\bf 100} (2012), pp. 161-170.

\bibitem{CK} B. Chow and D. Knopf, ``The Ricci Flow : An Introduction", Mathematical Surveys and Monographs Vol. 110, A.M.S., Providence, R.H., 2004.

\bibitem{Ca} Renate Carlson, ``$N$-ary algebras", Nagoya Math. J.{ \bf 78} (1980), pp. 45-56.

\bibitem{Ku} A.G. Kurosh, ``Multioperator Rings and Algebras", Russian Math. Surv. {\bf 24} (1) (1969), pp.1-13.

\bibitem{Kr} Yakov Krasnov, ``Differential Equations in Algebras", in Hypercomplex Analysis Trends in Mathematics, pp.187-205, Birkh\"auser, Boston, 2009.

\bibitem{Rod} A. Rogers, ``Supermanifolds : Theory and Applications", World Scientific Publ., River Edge, N.J., 2007.

\bibitem{PZ} A.D. Polyanin and V.F. Zaitsev, ``Handbook of Exact Solutions for Ordinary Differential Equations", 2nd edition, Chapman \& Hall / CRC, Boca Raton, U.S.A., 2003.\newline (See also http://eqworld.ipmnet.ru/index.htm)

\bibitem{PZS} D.E. Panayotounakos, Th. I. Zarmpoutis and P. Sotiropoulos, ``The General Solutions of the Normal Abel's Type Nonlinear ODE of the Second Kind", IAENG International Journal of Applied Mathematics {\bf 43} (3) (2013), pp.94-98.

\bibitem{In} E.L. Ince, ``Ordinary Differential Equations", Dover Publications Inc., N.Y., 1956, p.297.

\bibitem{CGM} J.F. Cari\~nena, J. Grabowski, and G. Marmo, ``Quantum Bi-Hamiltonian Systems", Int. J. Mod. Phys. A, {\bf 15} (2000), pp. 4797-4810.

\bibitem{OS1} A. Odesskii and V.V. Sokolov, ``Integrable matrix equations related to pairs of compatible associative algebras", J. Phys. A : Math. Gen. {\bf 39} (2006), 12447-12456.

\bibitem{OS2} A. Odesskii and V. Sokolov, ``Algebraic structures connected with pairs of compatible associative algebras", Int. Math. Res. Not. 2006, 43734 (2006) (arxiv : math/0512499).

\bibitem{Va} V.S. Varadarajan, ``Supersymmetry for Mathematicians : An Introduction", Courant Lecture Notes in Mathematics {\bf 11}, A.M.S., Providence, R.I., 2004.

\bibitem{BD} Bryce De Witt, ``Supermanifolds", Cambridge University Press, Cambridge, U.K., 1987.

\bibitem{Sc} Richard D. Schafer, ``An Introduction to Nonassociative Algebras", Academic Press, N.Y., 1966.

\bibitem{Ber} F.A. Berezin, ``Introduction to Superanalysis", ed. : A.A. Kirillov, D. Reidel Publishing Co., Dordrecht, Holland, 1987.

\bibitem{MS1} A.V. Mikhailov and V.V. Sokolov, ``Integrable ODEs on Associative Algebras", Commun. Math. Phys. {\bf 211} (2000), pp. 231-251.

\bibitem{MS2} A.V. Mikhailov and V.V. Sokolov, ``Symmetries of Differential Equations and the Problem of Integrability", in Lect. Notes in Phys. {\bf 767} (2009), pp.19-88.

\bibitem{Ol} P.J. Olver, ``Applications of Lie Groups to Differential Equations" Graduate Texts in Mathematics 107, Springer-Verlag, N.Y., 1986.

\bibitem{Mi} R.E. Mickens ed.,``Advances in the Applications of Nonstandard and Finite Difference Schemes", World Scientific Publ., Singapore, 2005.

\bibitem{BS} A.I. Bobenko and Yu. B. Suris, ``Discrete Differential Geometry : Integrable Structure", Grad. Studies in Math. Vol. {\bf 98}, Providence, R.I., A.M.S. 2008.

\bibitem{GRSW} B. Grammaticos, A. Ramani, J. Satsuma, and R. Willox, ``Discretizing the Painlev\'e equations \`a la Hirota-Mickens", J. Math. Phys. {\bf 53} (2012), 023506, 24 pp..

\bibitem{Re} W.T. Reid, ``A matrix differential equation of Riccati type", Am. J. Math. {\bf 68} (1946), pp.237-246.

\bibitem{Le} J.J. Levin, ``On the matrix Riccati equation", Proc. A.M.S. {\bf 10} (1959), pp. 519-524.

\bibitem{PW} A.V. Penskoi and P. Winternitz, ``Discrete matrix Riccati equations with superposition formulas", J. Math. Anal. Appl. {\bf 294} (2004), pp.533-547.

\bibitem{RSWa} R.S. Ward, ``Discretization of integrable systems", Phys. Lett. A {\bf 165} (1992), pp. 325-329.

\bibitem{WZG} Wu Ke, Zhao Weizhong, and Guo Hanying, ``Difference discrete connection and curvature on cubic lattice", Science in China, Series A : Mathematics {\bf 49} (2006), pp. 1458-1476.

\bibitem{DMH} A. Dimakis and F. M\"uller-Hoissen, ``Solutions of matrix NLS systems and their discretizations : a unified treatment", Inverse Problems {\bf 26} (2010), 095007, 55 pp..

\bibitem{Hie} J. Hietarinta, ``Definitions and Predictions of Integrability for Difference Equations", in Symmetries and Integrability of Difference Equations, eds : D. Levi, P. Olver, Z. Thomova and P. Winternitz, London Mathematical Society Lecture Note Series Vol. 381, Cambridge U. Press, Cambridge, U.K., 2011, pp.83-114.

\bibitem{Ha} Mike Hay, ``A Completeness Study on Certain $2 \times 2$ Lax Pairs Including Zero Terms", SIGMA {\bf 7} (2011), 089, 12pp..

\bibitem{Ben} Bennett Chow, Sun-Chin Chu, David Glickenstein, Christine Guenther, James Isenberg, Tom Ivey, Dan Knopf, Peng Lu, Feng Luo, and Lei Ni, ``The Ricci Flow : Techniques and Applications : Part I : Geometric Aspects", Mathematical Surveys and Monographs Vol. 135, A.M.S., Providence, R.I., 2007.

\bibitem{MMAGY} Warner A. Miller, Jonathan R. McDonald, Paul M. Alsing, David Gu, Shing-Tung Yau, ``Simplicial Ricci Flow", arxiv : 1302.0804.

\bibitem{HK} N.C. Hopkins and M.K. Kinyon, ``Quadratic Differential Equations in $\Z_2$-Graded Algebras", Trans. of Am. Math. Soc. {\bf 351} (1999), pp. 4545-4559.  


\end{thebibliography}
\end{document}